\documentclass[drafts]{agujournal}
\journalname{Journal XXX}
\pdfoutput=1
\usepackage{amsmath,amsfonts,amssymb,color}
\usepackage{graphicx}
\usepackage{hyperref}
\usepackage{subfig}
\usepackage{color}
\usepackage{times}
\usepackage{float}
\usepackage[mathscr]{euscript}
\usepackage{setspace}
\usepackage{caption}

\floatstyle{boxed}
\restylefloat{figure}
\newcommand{\eps}{\epsilon}

\begin{document}

\title{The Mid Pleistocene Transition from a Budyko-Sellers Type Energy Balance Model}
\authors{Esther R. Widiasih, Malte F. Stuecker, Somyi Baek \affil{1,2,3}}

\begin{center}
\today

\end{center}


\affiliation{1}{Mathematics and Science Division, University of Hawaii - West Oahu}
\affiliation{2}{Center for Climate Physics, Institute for Basic Science, Pusan National University}
\affiliation{3}{School of Mathematics, University of Minnesota - Twin Cities}

\correspondingauthor{Esther R. Widiasih}{widiasih@hawaii.edu}


\begin{abstract} 
A conceptual model of the Plio-Pleistocene glacial cycles is developed based on the Budyko-Sellers type energy balance model. The model is shown to admit a phenomenon like the Mid-Pleistocene transition, capturing the essence of the albedo and the temperature precipitation feedbacks.
\end{abstract}

\section{Introduction}
\subsection{General problem of glacial cycles and the MPT}
In the past 800 kyr, the Earth's climate has experienced interglacial-glacial cycles (or simply glacial cycles from here on) with a 100 kyr  periodicity. The Milankovitch theory proposes that the glacial cycles is driven by a linear response to the summer insolation signal at high northern latitude. However, while such signal possesses a weak 100 kyr period of the Earth's orbital eccentricity, it is dominated by the obliquity signal with a period of 41 kyr. This problem is also known as the 100 kyr problem. 

An even more fasinating problem is the change in the periodicity of the glacial cycles, that occur about 1.2 - 0.5 Mya, known as the Mid Pleistocene Transition (MPT). Prior to the MPT, the glacial cycles occur at a shorter period of 41 kyr, having a lower amplitude. After the MPT, each cycle in the glacial period lengthens to 100 kyr, having a higher amplitude and a more asymmetrical feature resembling a saw-tooth. Both the 100 kyr and the MPT problems call for a non-linear treatment in the modeling of the glacial cycles, having a parameter shift that induces another shift in the response variables.

By design, conceptual models are highly simplified representation of the planetary climate, because these models are tools to probe big picture ideas. Numerous conceptual models have found success in exhibiting differing mechanisms for both the 100 kyr  glacial cycles and the MPT. A few well studied examples are \cite{paillard2004antarctic}, \cite{maasch1990low}, and \cite{tziperman2003mid}. In all of these examples, the state variables are global averages. The model proposed here is based on what's known as the Budyko-Sellers model (for examples, see the discussions in \cite{north1975theory, north1981energy, sellersglobal, tung2007topics}), a latitudinally averaged energy balanced model. The one used in particular is the one first written by M. Budyko \cite{budyko1969effect} to model the annual temperature distribution over hemisphere, assuming a symmetric water planet. In using such model, one may benefit by exploiting a balance principle on the  latitudinal position of some significant features, such as the glacier's edge and the snowline. Indeed, the glacier's edge and the snowline positions are two of the state variables in the proposed glacial model shall represent, and the third state variable approximates the planetary average temperature. The model admits sustained glacial cycles (free oscillations) without any external forcing from orbital elements. The details of the modeling process, the limit cycle and other mathematical properties of the model are discussed in \cite{wwhm2016periodic}.

\subsection{Background of the Glacial Flip Flop Model}
Because of the switching feature of this model, this model is called the \emph{glacial flip flop} model, as it flip and flops between glaciating and de-glaciating states. The idea of a multi-state climate model is not new. For example, in the conceptual model proposed by \cite{paillard1998timing}, three climate regimes: i (interglacial), g (mild glacial), G (full glacial). The global circulation model explored in \cite{abe2013insolation} shows two distinct climate regimes, one for a positive and the other for negative total mass balance state. These two models are clearly at the opposite ends of the spectrum in terms of model complexity, and yet, they both show transient evolution of the climate, weaving in and out of these  climate regimes. 

Each multi-state climate model of the glacial cycles discussed above is equipped with a switch mechanism. What might be the driver of the switch? A compelling discussion in \cite{raymo2006plio} (par. 15) posed a conjecture that what determines the maximum ice sheet size is perhaps the albedo change causing  positive and negative feedbacks affecting the temperature-precipitation feedback. The change in the interplays between the two feedbacks might also be the difference between the early and late Pleistocene. The 3 state variables in the  \emph{glacial flip flop} model captures exactly these interplays as the albedo feedback affects temperature and ice sheet edge maximum latitudinal position. The switching mechanism of the glacial flip flop is governed by the energy balance principle as was done in the Budyko-Sellers type models, and the mass balance principle, in the same spirit as the global circulation model in \cite{abe2013insolation}. 

The glacial flip flop model featured here may be reminiscent of the conceptual model of the ice sheet proposed in \cite{weertman1976milankovitch, kallen1979free}, featuring the \emph{precipitation and temperature feedback}. There are many similarities between the two models: the use of Budyko-Sellers type energy balance model to govern the temperature, the parameterization of the latitudinal extent of the some ice cover, and finally, the use of mass balance (accumulation-ablation) to drive the ice volume evolution. The difference, however, is apparent to distinguish the two models, with the main difference being the existence of the parametrization of the snow line in the glacial flip flop model. 

Two prior works staged the foundation of the glacial flip flop model. In the first work \cite{widiasih2013dynamics}, a parametrization of the albedo line in an aquaplanet is incorporated into the existing Budyko energy balance model of the spatially dependent temperature distribution. The albedo line is thought to represent a latitudinal line separating perenially snow covered region from that where the snow melts in the summer, akin to a climactic snow line. Indeed, this is in contrast to most of the Budyko-Sellers type of energy balance model where the albedo function is dependent on the temperature at that latitude. As the albedo feedback depends on the area of differing surface albedo, the latitudinally dependent albedo function holds the advantage in representing the yearly average energy absorbed/ reflected by the system. Assuming a spherical planet, when the spatial dependence is represented as the sine of the latitude (as was done in \cite{tung2007topics}), a snow cap extending from the pole to the snow line (or albedo line) at latitude $\theta$ covers exactly 1-$sin$($\theta$) of the planet surface area having higher albedo, while $sin$($\theta$) is the surface area with lower albedo. Therefore, the incorporation of the surface albedo feedback could be done more directly.

While the zonally averaged Budyko energy balance model of the temperature distribution is often referred to as one dimensional energy balance model because its spatial dependent variable is one dimensional (see eg. \cite{mcguffie2005climate}), mathematically speaking, it is an infinite dimensional system of integro-partial differential equation. The second work staging the foundation of the glacial flip flop model was presented in \cite{mcgwid2014simplification}, and showed the reduction of that snow line-zonal temperature distribution into a manageable system of ordinary differential equations of two variables. Because of this reduction, the temperature distribution  could now be represented by one variable, the evolution of which captures the surface albedo feedback parametrized in the snow line, acting as a line separating lower and higher albedo surface.

The addition of the third variable in the glacial flip flop model is done to capture the mass balance interplay between the accumulation zone and the ablation zone. The albedo line of the second work is driven by the temperature at the snow line that controls the maintenance of perenial snow cover, and thereby, captures the essence of \emph{precipitation and temperature feedback}. The third variable, representing the edge of the ice sheet, models the evolution of the ice extent by capturing the mass balance between the accumulation zone (poleward of the snow line) and the ablation zone (between the snow line and the ice sheet extent). As one considers the thousand years time scale of the glacial cycles, the evolution of the annual average temperature should be considered working at a relatively faster rate, while the snow line and ice line evolution should be at relatively slower rates  (see the discussion in \cite{mcgwid2014simplification}). This modeling approach allows an exploration into the conjecture on what determine the maximum ice sheet as posed by \cite{raymo2006plio}.

The evolution of the snow line (or albedo line) is driven by the temperature difference between the snowline temperature and a critical temperature. When the snowline temperature is warmer than a critical temperature, the snowline retreats toward the pole, otherwise, it advances toward the equator. The modeling approach incorporating a critical ice-forming temperature has also been utilized by others, for example in \cite{tziperman2003mid}. In that work, the ice-forming temperature is varied slowly from -13$^o$C 1.5 Ma BP to -3$^o$C at 500 kyr BP in order to simulate the deep ocean cooling. The recent work by \cite{tzedakis2017simple} presented a statistical model correctly  predicting  every complete deglaciation over the past 1 Myr. This statistical model incorporates the observations that the energy threshold to start deglaciation has increased in the past 1 Myr and that the increase in the energy threshold causes \emph{skipped deglaciation}, which in turn results in larger ice sheet and longer glaciation. The shift in the ice-forming critical temperature modeling approach done by \cite{tziperman2003mid} is certainly consistent with the more recent statistical finding. As the discussion of this model shows in the next section, the same shift in the ice-forming critical temperature parameter drives a bifurcation, which in turn causes an increase in the amplitude and the period of the glacial cycles. 

Here, we pose the main question of this work: given that the incoming solar radiation forcing is a function of eccentricity and obliquity, could the glacial flip flop model simulate the transition from 41 kyr to 100 kyr glacial cycle period? Could the model simulate the transition even if the eccentricity is held constant? In the following sections, we discuss the details of the glacial flip flop model, some mathematical  interpretations of the model, and some simulations showing the glacial cycles and the transition from the 41kyr to 100kyr period.

\section{Model description}
The glacial flip flop model assumes the existence of two regimes: glaciation and inter-glaciation, and the system flips and flops between the two regimes. The distinguishing features of the two regimes lies on two parameters: the ice forming critical temperature and the ablation rate. The ice forming critical temperature of glaciation regime is higher than that of the inter-glaciation regime. The three model variables, $w, \eta \, \text{, and }\xi$, respectively simulate the global annual temperature, the sine of the latitude of the snow/ albedo line, and the sine of the latitude of the maximum ice cover extent (ice extent from here on). Hence, variables $w$ are real valued, while $\eta$ and $\xi$ live on the unit interval $[0,1]$, with $0$ representing the equator and $1$ the pole. Below we briefly present the formulation of the glacial flip flop model, a detailed explanation of the modeling can be found in \cite{wwhm2016periodic}. 

The evolution of the temperature $w$, derived from the Budyko energy balance model, reflects the ice albedo feedback, and so, its equation of motion is coupled with the albedo line $\eta$. The motion of the albedo line $\eta$ reflects the precipitation and temperature feedback, hence, its evolution equation is coupled with temperature $w$ and it is dependent on the ice forming critical temperature. The evolution of the ice extent $\xi$ relies on the mass balance principle, hence its equation of motion keeps track of the difference between ablation rate and accumulation rate. Ablation happens on the zone between $\xi$ and $\eta$, while accumulation takes place between $\eta$ and the pole, $1$. The evolution of the maximum ice extent $\xi$ depends linearly on the size of the ablation zone $\eta-\xi$ and the accumulation zone $1-\eta$.

\subsection{The variables and equations}

During a glacial period, that is when accumulation exceeds ablation, the ice extent is advancing. The equations of motion for $w, \eta, \text {and } \xi$ are: 

\begin{subequations}\label{adv}
\begin{align}
\dot{w} &= -\tau ( w-F(\eta) ) = f^a_1(w,\eta, \xi ) \label{advA}\\
\dot{\eta}&= \rho ( w-G_a(\eta,Tc_a) )= f^a_2(w,\eta,\xi) \label{advB}\\
\dot{\xi} &= \eps \left( b_a(\eta-\xi)-a(1-\eta) \right)= f^a_3(w,\eta,\xi). \label{advC}
\end{align}
\end{subequations}

During an inter-glacial period, that is when the ice extent is retreating, the equations of motion for $w, \eta, \text {and } \xi$ are:

\begin{subequations}\label{ret}
\begin{align}
\dot{w} &= -\tau ( w-F(\eta) ) = f^r_1(w,\eta, \xi ) \label{retA}\\
\dot{\eta}&= \rho ( w-G_r(\eta,Tc_r) )= f^r_2(w,\eta,\xi) \label{retB}\\
\dot{\xi} &= \eps \left( b_r(\eta-\xi)-a(1-\eta) \right)= f^r_3(w,\eta,\xi). \label{retC}
\end{align}
\end{subequations}

The two regimes have two distinct  ice forming critical temperature $Tc_a \text { and } Tc_r$ as well as  ablation rates $b_a \text{ and } b_r$. The value of the interglacial ice forming critical temperature as the ice extent is retreating is set to  $Tc_r \approx -10^o$C, and is close to that of today's climate. This value is used eg. in \cite{tung2007topics}. The interglacial period ice forming temperature $Tc_r$ is lower than $Tc_a \approx -6^o \text{ to } -10^o$C. Note that the range of both the ice forming critical temperatures are well within that used by \cite{tziperman2003mid}. The ablation rate or advancing ice sheet, $b_a$ is lower than $b_r$, the rate when the ice extent is retreating. The switch between the two regimes happens at the discontinuity plane, when $b (\eta - \xi) - a (1 - \xi)=0$, where $b$ is some critical ablation rate, having the value  $b_a < b < b_r$.

In this model, the orbital forcing  enters through the function $F(\eta)$ due to its dependence on the incoming solar radiation parameter $Q$ (which depends on the eccentricity) and the distribution parameter $s_2$ (which depends on obliquity). Since the variables in the flip flop model simulate annual averages,  eccentricity and obliquity dependent orbital elements can be included in the solar insolation function, leaving out any treatment on precession. The details of the functions $F, G_a, G_r,$ and the parameters used can be found in \textbf{TABLE 1}. 

\subsection{Some mathematical interpretations}
\begin{figure}[H]
\centering
\subfloat[From left to right, the phase portraits of $w-\eta-\xi$ of the autonomous system with  varying in equation \eqref{advA} the critical ice forming temperature $Tc_a = -9, -7, -5^oC$, keeping $Tc_r = -10^oC$. The mass balance is positive in the blue region, and negative in the red region. The diagonal plane along the $w$ axis, nearly orthogonal to the page is the discontinuity plane separating the glaciation and interglaciation regimes. The red and blue dots are the virtual equilibria of the system in equations \eqref{adv} and \eqref{ret}] 
{\includegraphics[width=0.3\textwidth]{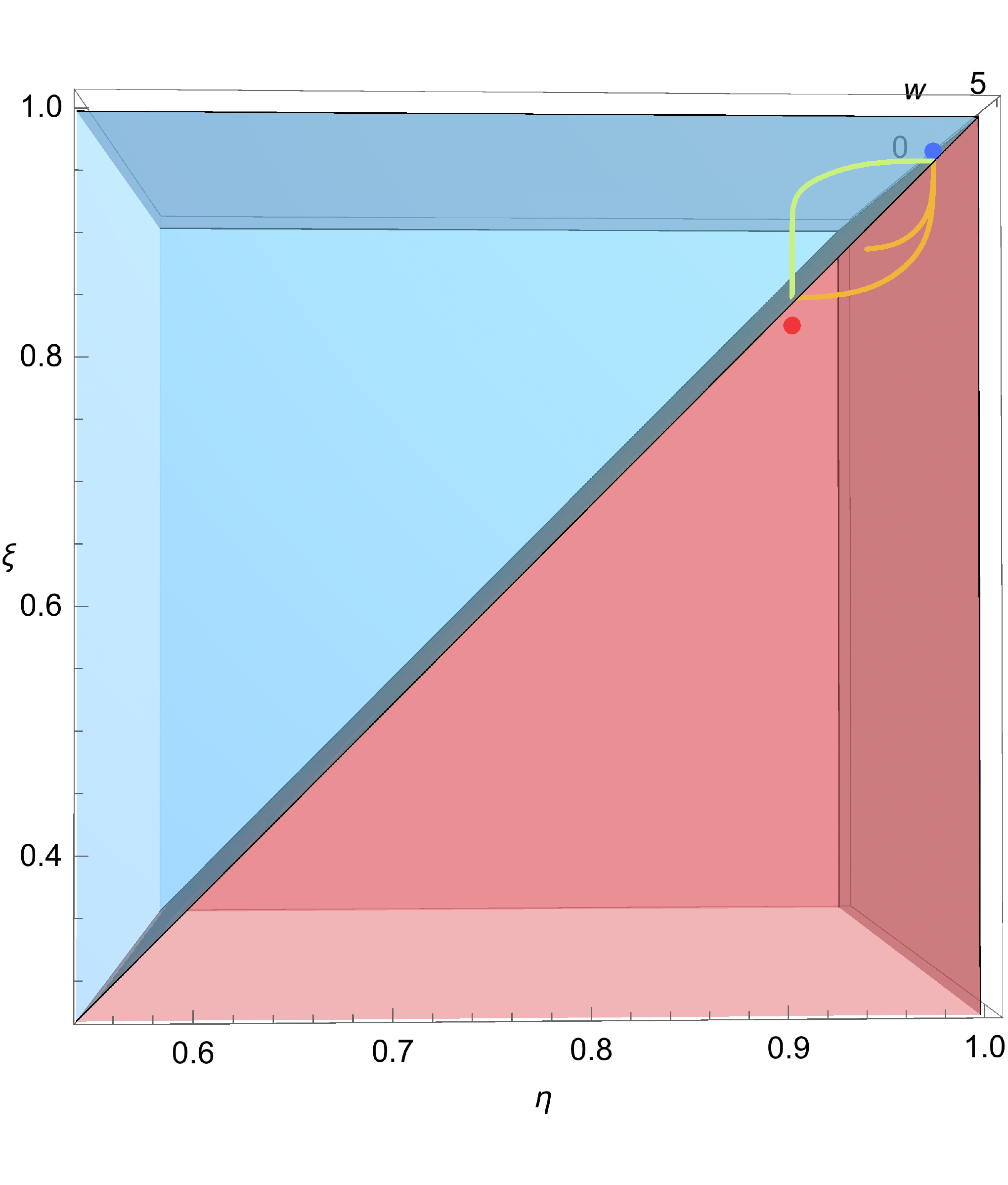} 
\includegraphics[width=0.3\textwidth]{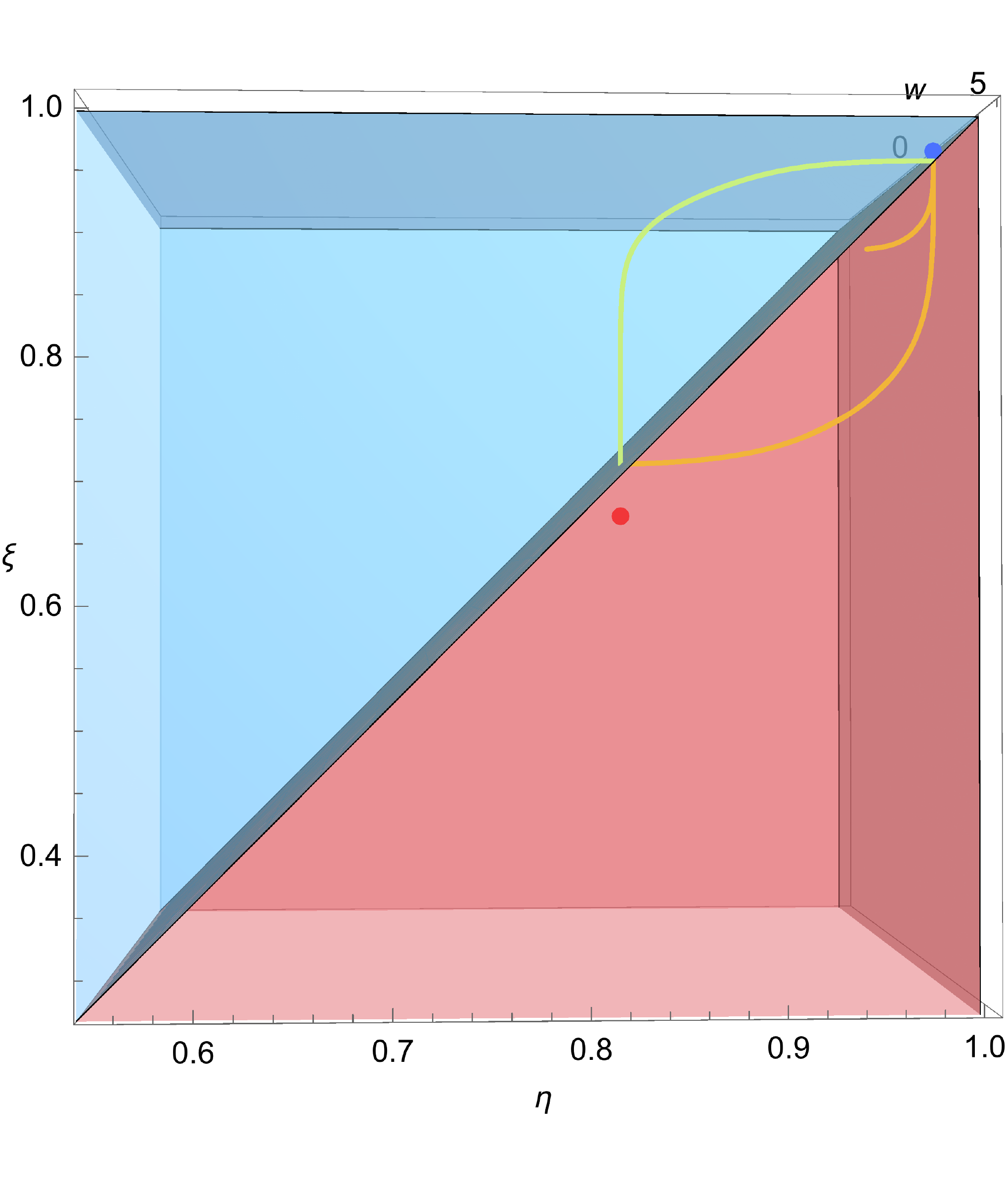} 
\includegraphics[width=0.3\textwidth]{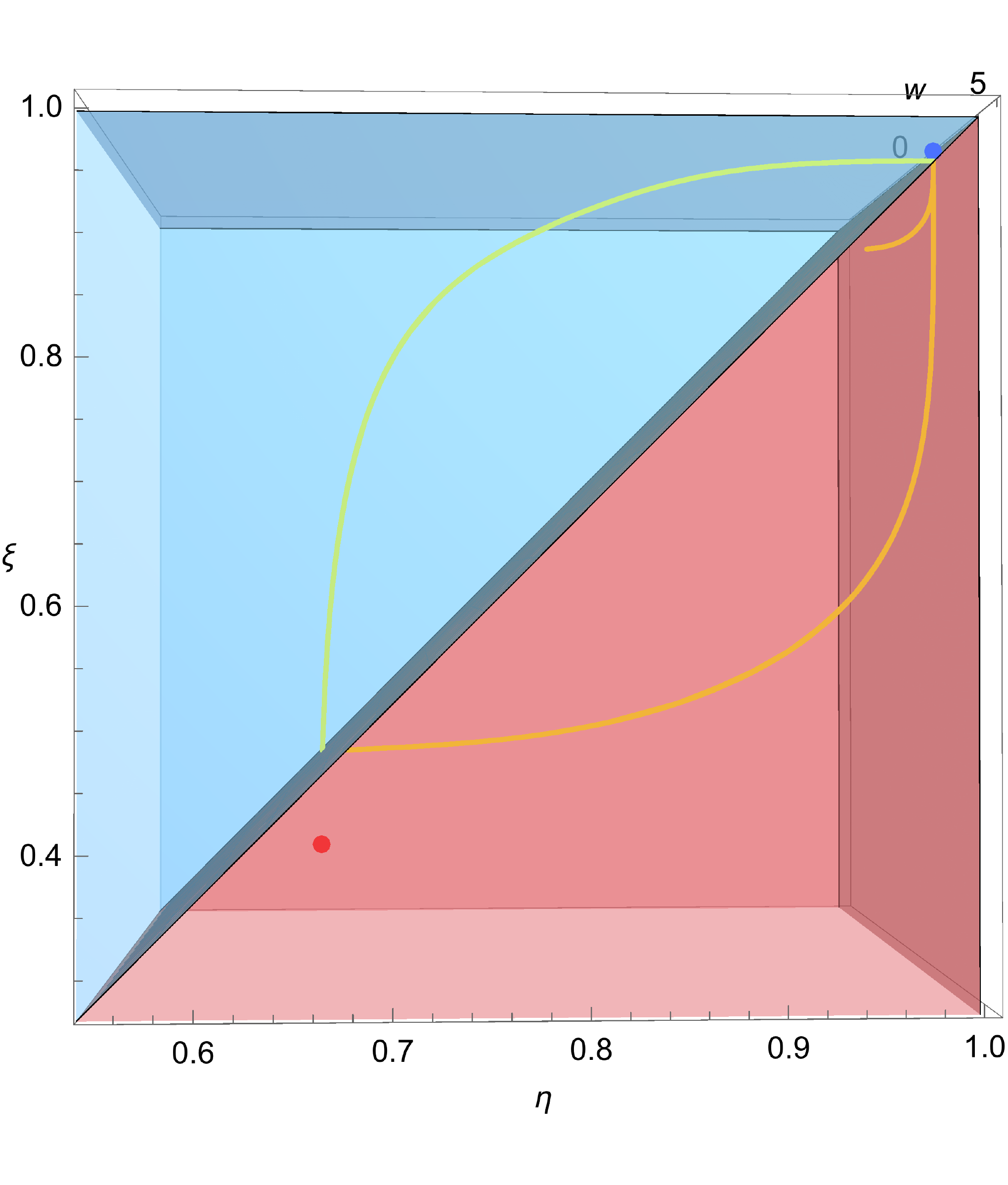} 
}

\subfloat[The corresponding global mean temperature evolution for $Tc_a = -9, -, -5^o$C. ] 
{\includegraphics[width=0.3\textwidth]{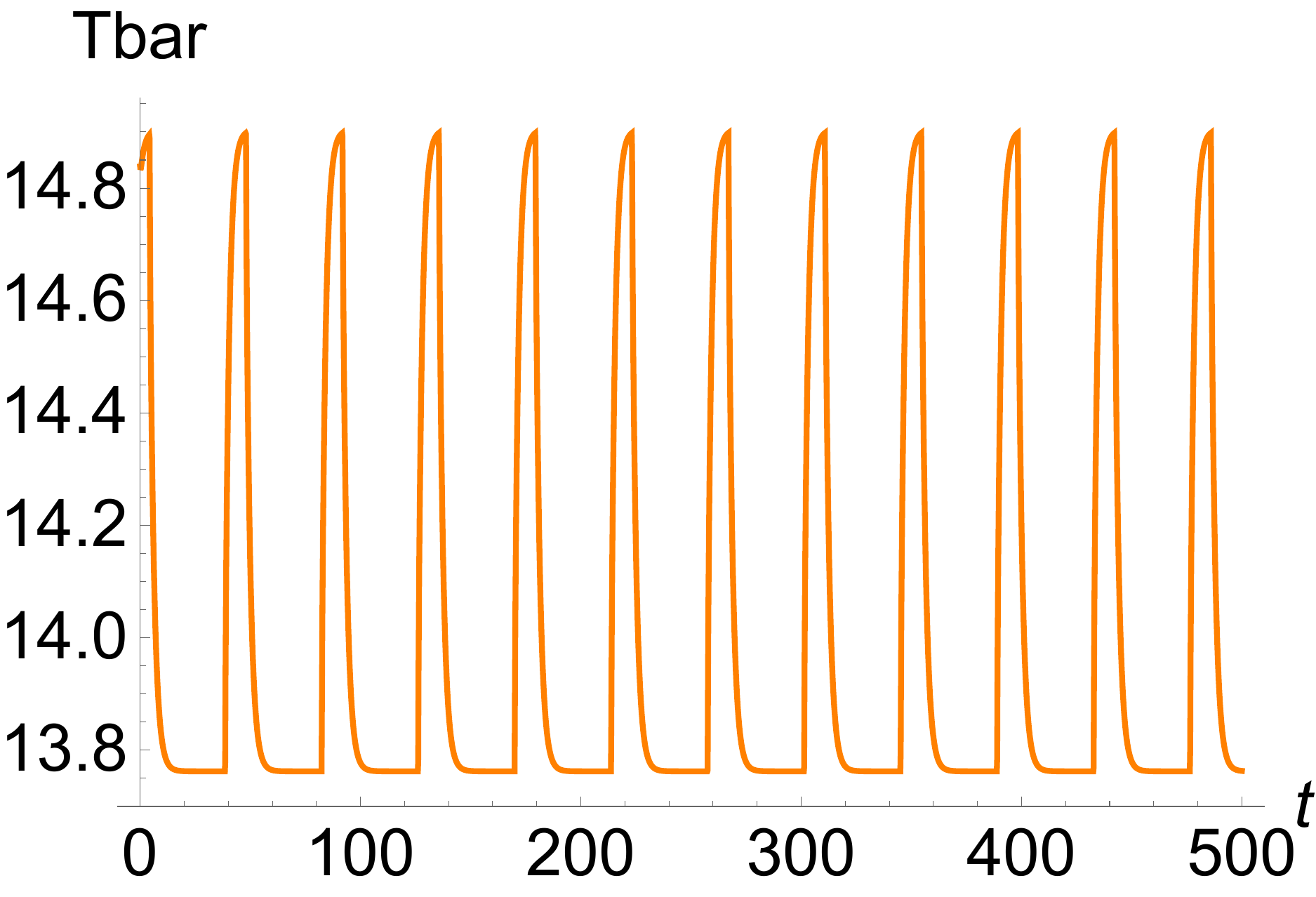} 
\includegraphics[width=0.3\textwidth]{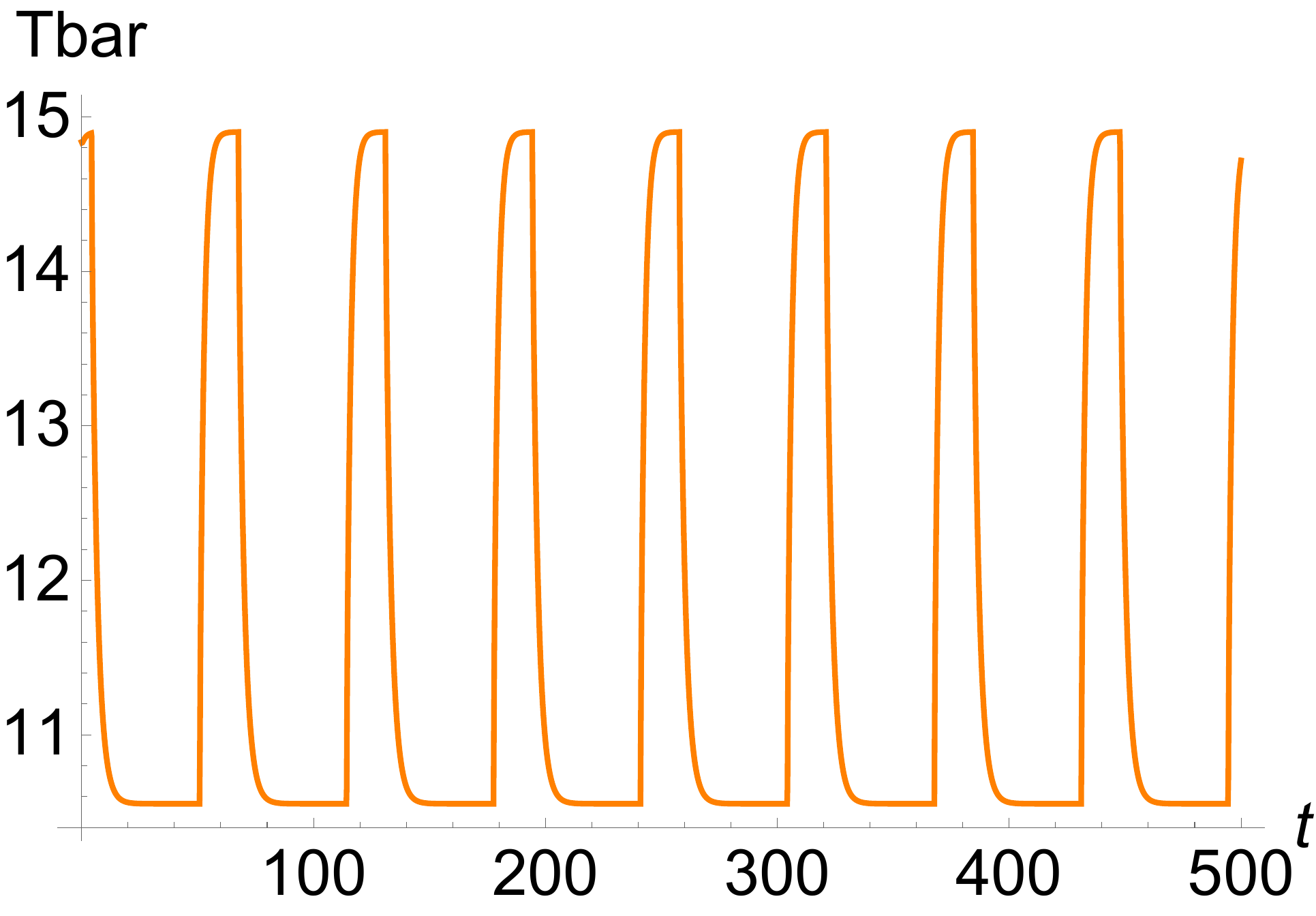} 
\includegraphics[width=0.3\textwidth]{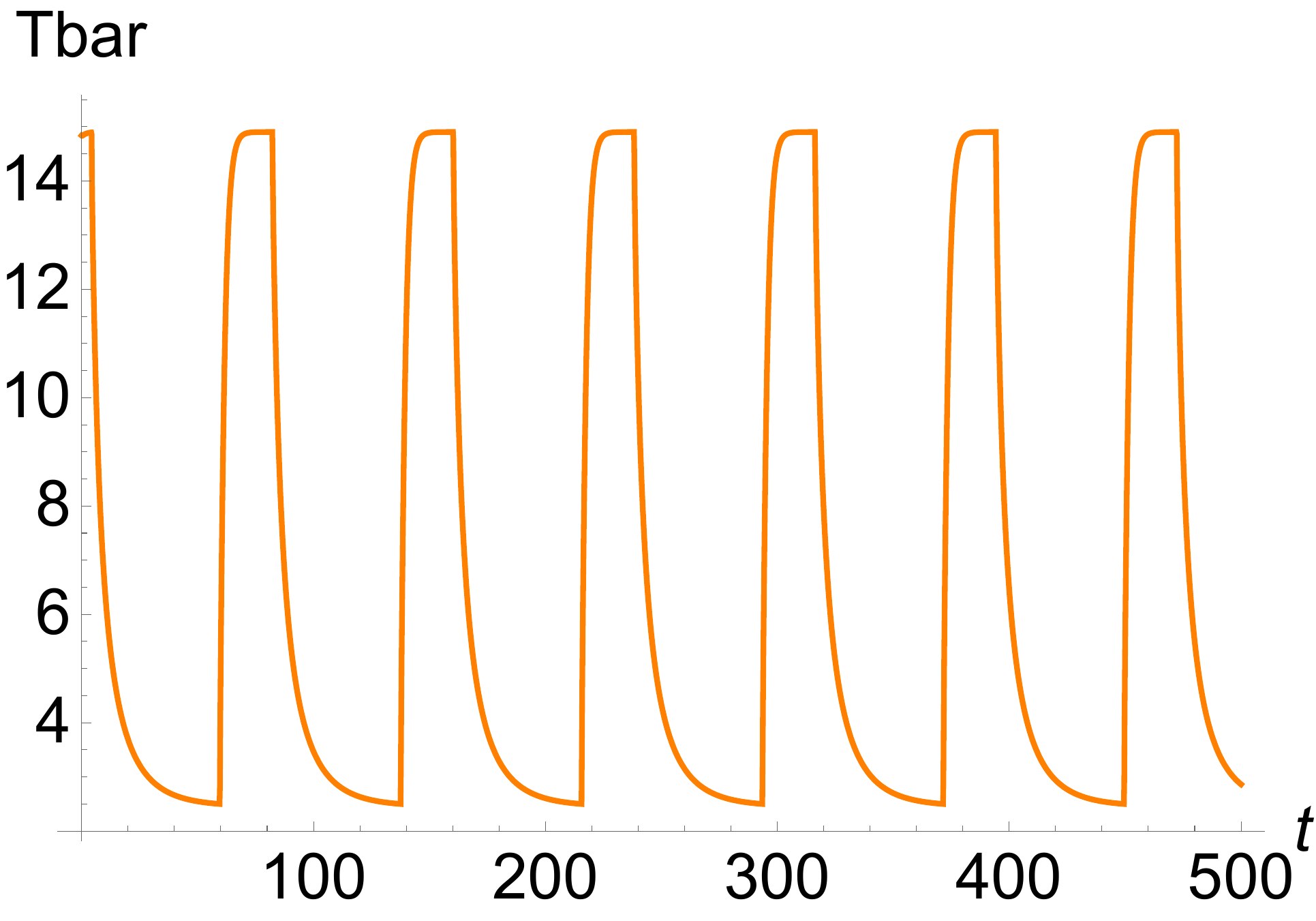} 
}

\subfloat[The corresponding ice extent$\xi$ and snow line $\eta$ evolution for $Tc_a = -9, -, -5^o$C.  ] 
{\includegraphics[width=0.3\textwidth]{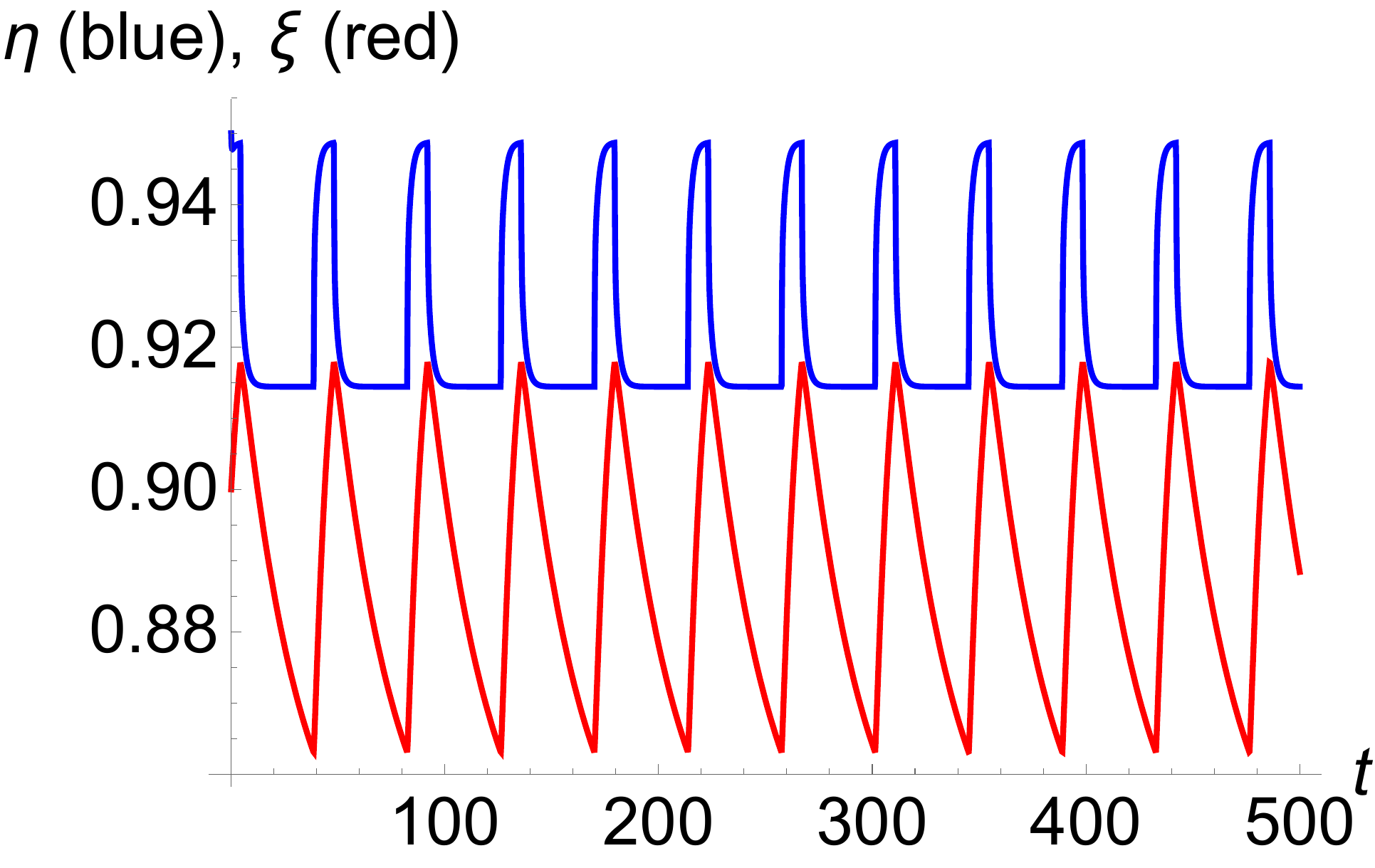} 
\includegraphics[width=0.3\textwidth]{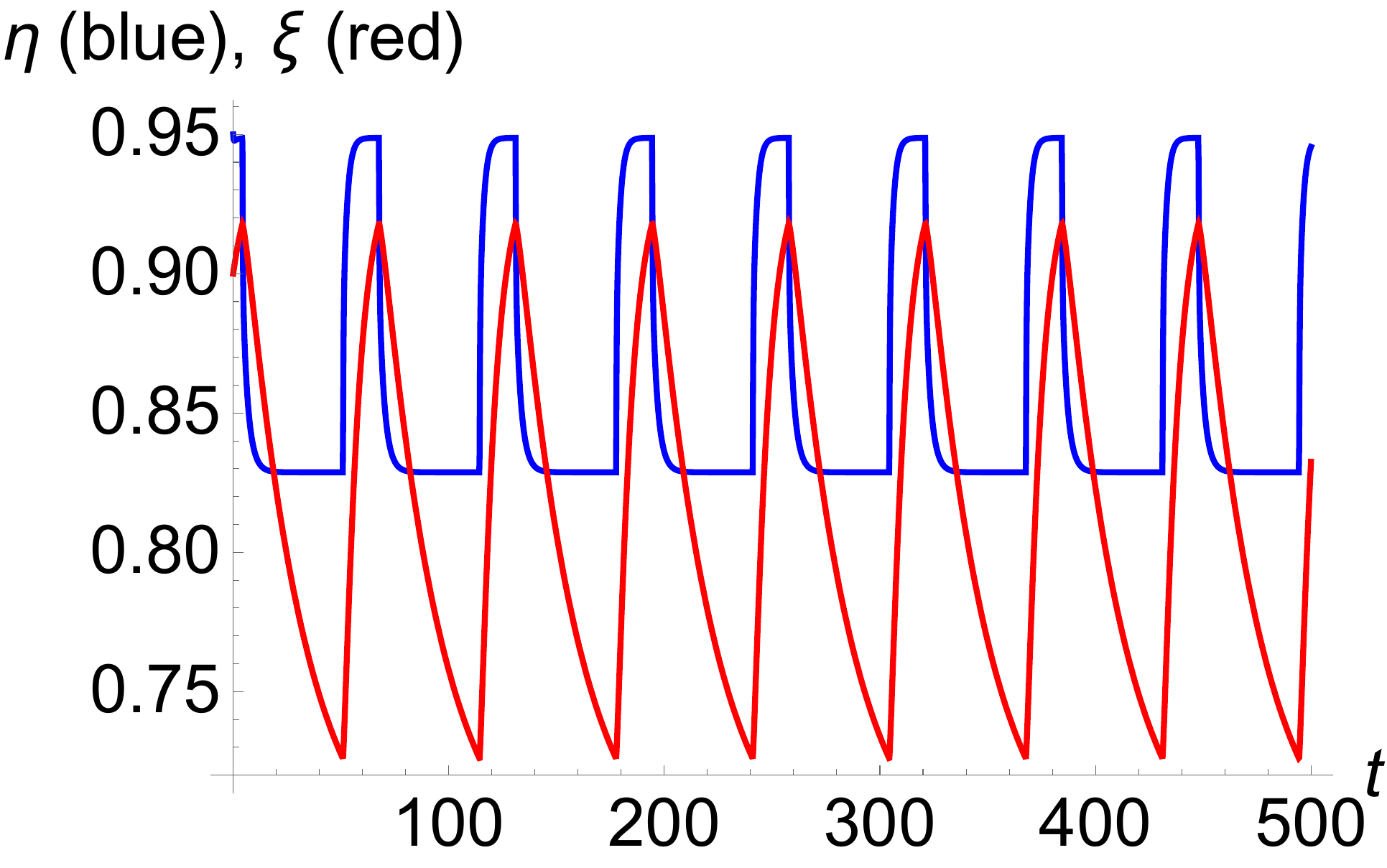} 
\includegraphics[width=0.3\textwidth]{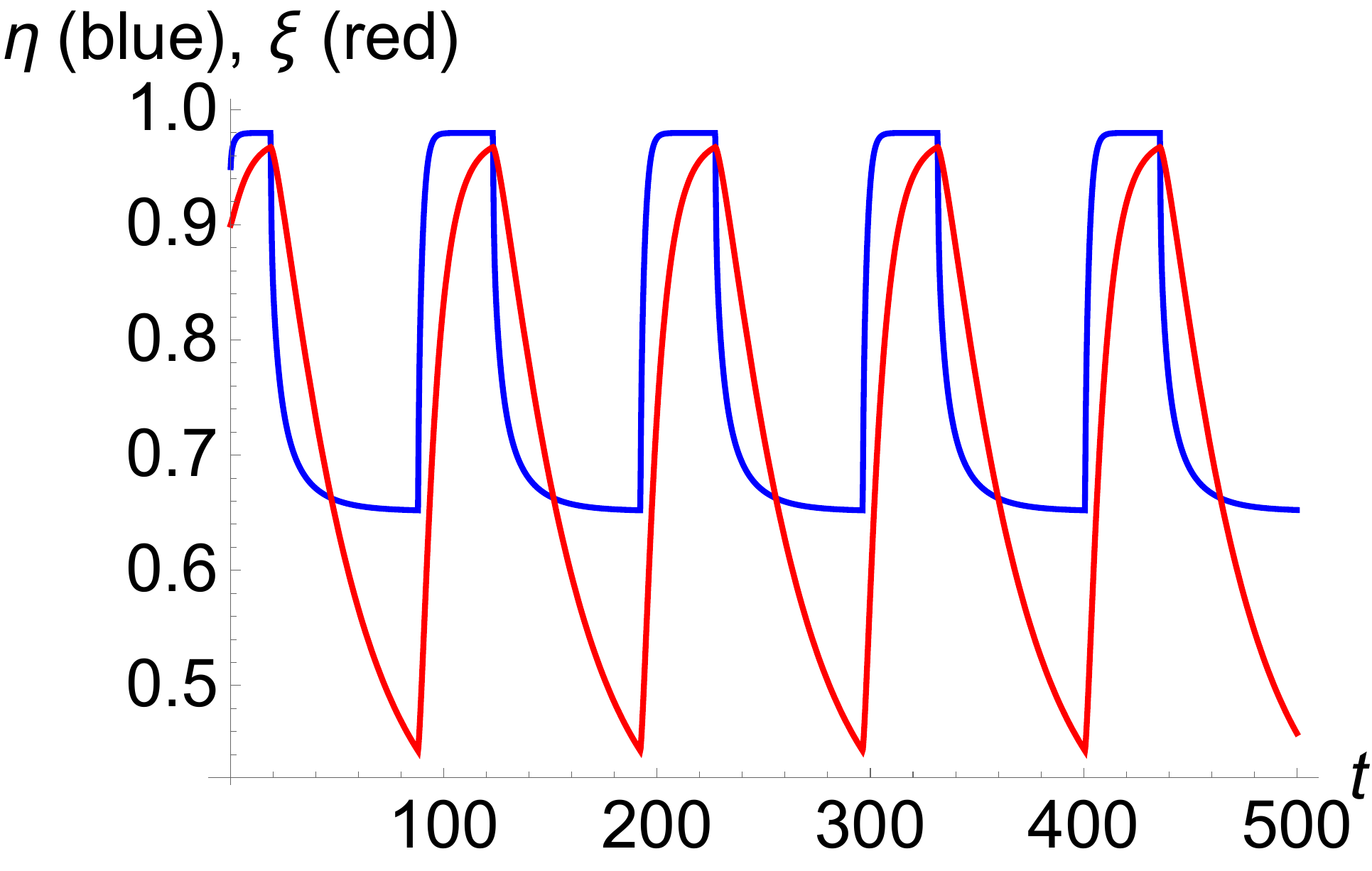} 
}
\caption{The behavior of the glacial flip flop system as the ice forming critical temperature $Tc_a$ is varied. Keeping $Tc_r$ constant, the amplitude and the period of the cycle grow as $Tc_a$ increases. } \label{fig1}
\end{figure}

Within some reasonable parameter regimes, the glacial flip flop model is able to generate a cycle with varying amplitude and periodicity. In Figure \ref{fig1}, the only varying parameter is the ice forming critical temperature during the glaciation period $Tc_a$, which is increased from $-9^o$C to $-5^o$C. In this particular exploration, both solar orbital parameters affecting eccentricity $Q$ and obliquity $s_2$ are fixed constant. The amplitude and period of the cycle increases with $Tc_a$. This prompts a current work in \cite{makarenkov2018}  showing the  existence of a bifurcation of a limit cycles from a singularities of the nonsmooth system in equations \eqref{adv} and \eqref{ret}. 

The dynamics of the cycle is controlled by the two virtual equilibria (blue and red dots in the phase plane of Figure \ref{fig1}). The trajectory of the glaciating regime is trying to reach the virtual equilibrium associated with the glaciating regime (the blue dot), while that of the deglaciation regime is aiming for the interglacial virtual equilibrium (the red dot). In particular, notice that in the glacial flip flop model, the parameter $Tc_a$ controls the start of the deglaciation; when $Tc_a$ is closer to $Tc_r$, the deglaciation starts earlier than when $Tc_a$ is at a higher value. Hence, exploration in the direction of increasing ice forming threshold $Tc_a$ will be consistent with the previously cited work by  \cite{tzedakis2017simple} on the increasing threshold for deglaciation as well as with a similar modeling experiment done by \cite{tziperman2003mid}.

\begin{figure}[H]
\centering

\subfloat[The time dependent ramp functions for $Tc_a(t)$ (purple) and for $Tc_r(t)$ (pink), are used in this exploration (time $t = 0$ is 5 Myr). The resulting time series of the ice sheet extent variable $\xi$ (orange) shows growth in period and amplitude.]
{
\includegraphics[width=0.4\textwidth]{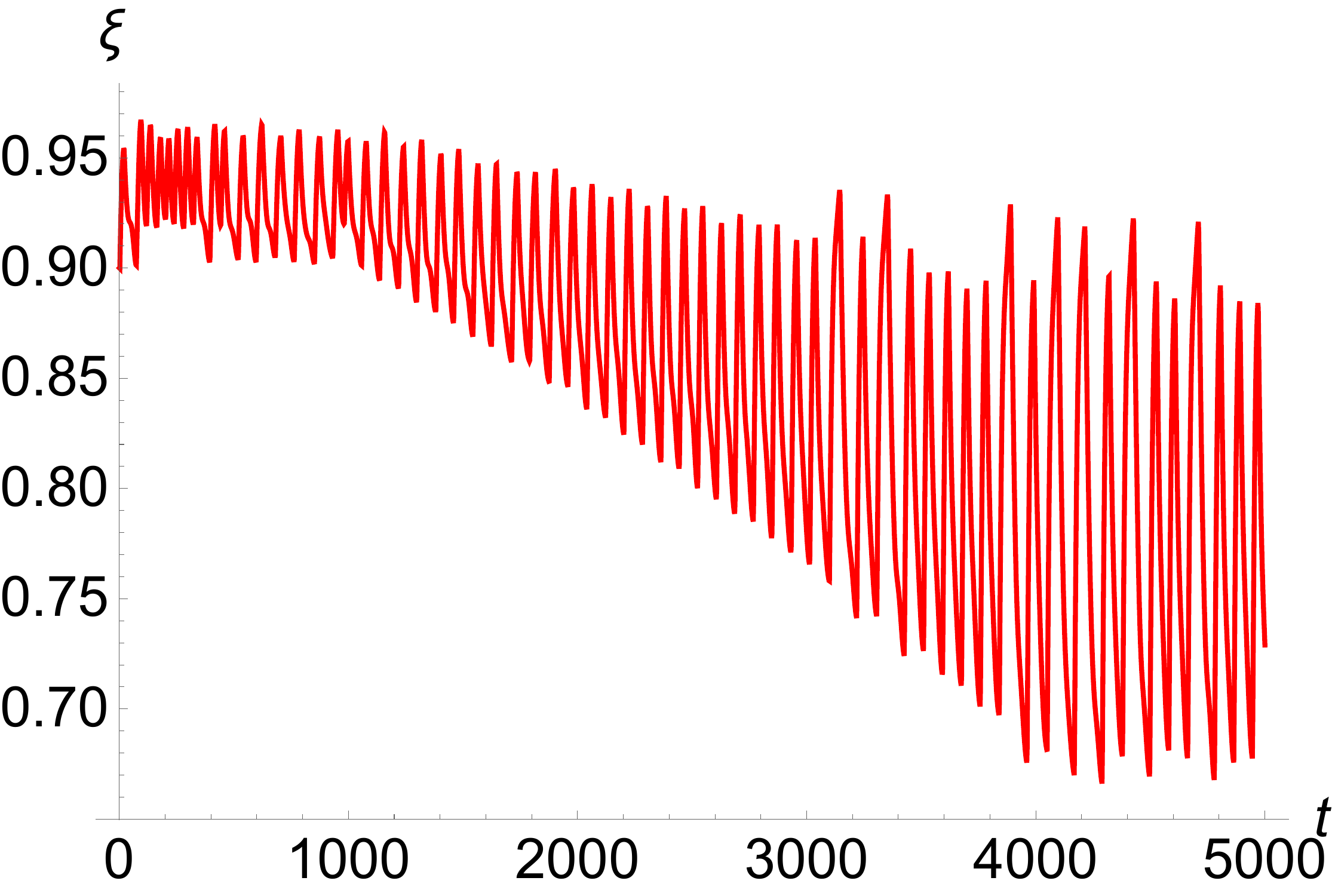}
\includegraphics[width=0.4\textwidth]{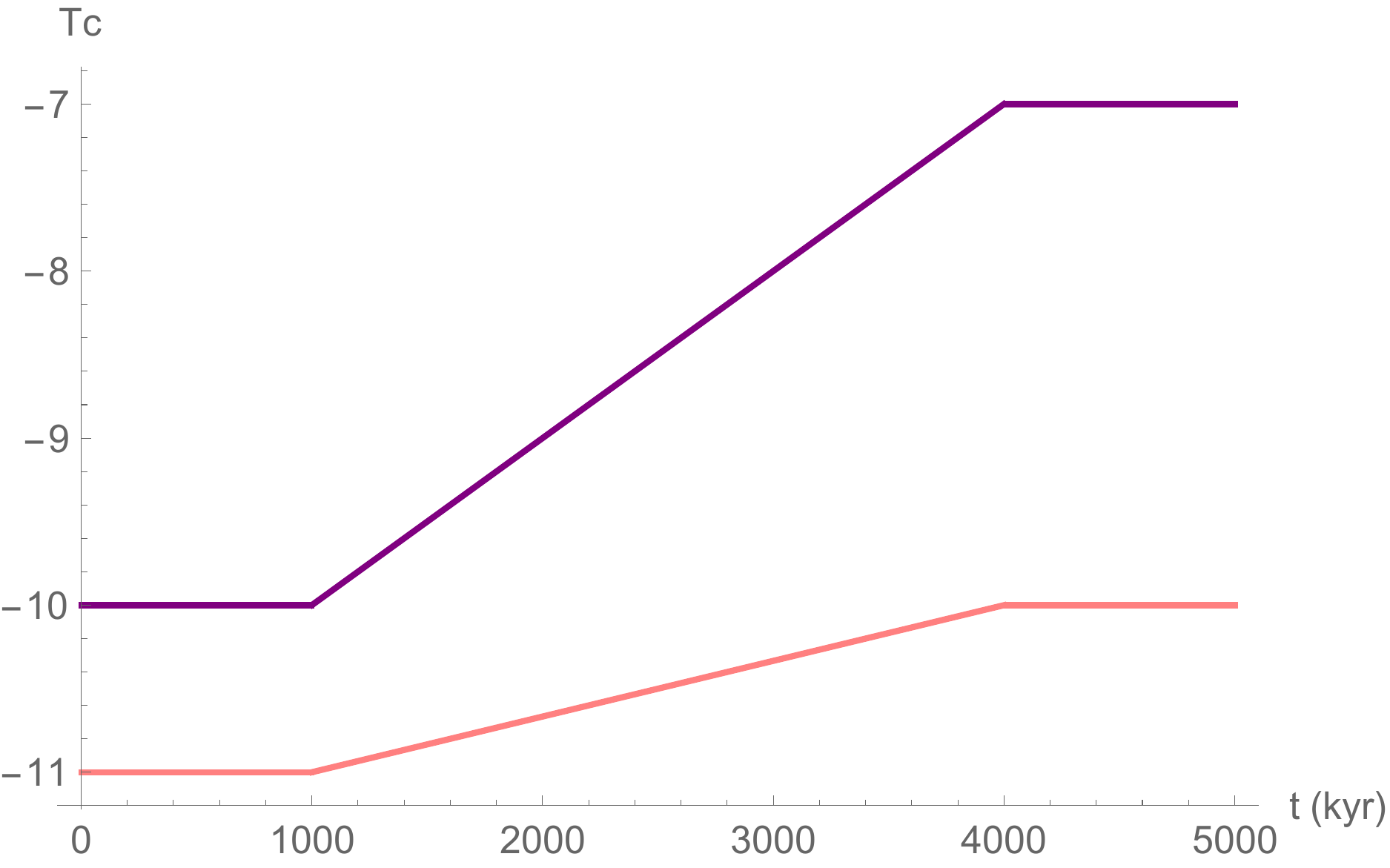}
}

\subfloat[The phase portrait of the glacial flip flop system with increasing ramp functions for $Tc_a$ and $Tc_r$. Trajectory in yellow is from 5Myr to 4Myr, orange is from 4 Myr to 1Myr, and red is the last 1 Myr.]
{
\includegraphics[width=0.8\textwidth]{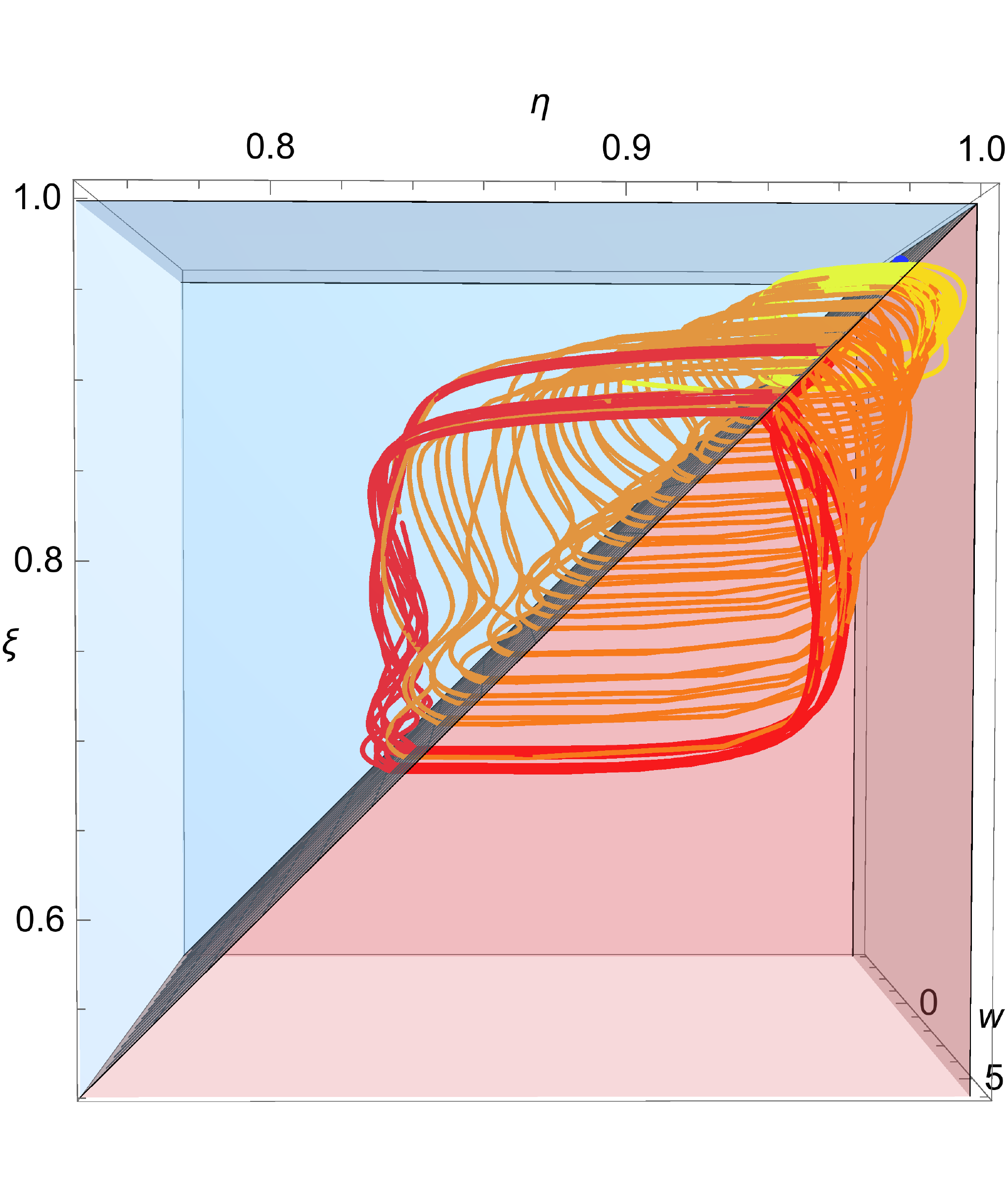}
}

\caption{Glacial flip flop model using time depenendt ramp functions.} \label{fig2}
\end{figure}

Drawing inspirations from this, two ramp functions for $Tc_a$ and $Tc_r$ increasing over the past 5 Myr may be used to capture a phenomenon like the Mid-Pleistocene Transition. Using ramp functions shown in Figure \ref{fig2}, phase portraits of the trajectory of the system as well as the resulting maximum ice extent $\xi$ appears to show a transition in the period and amplitude of the system. This particular exploration shows that the cycle in the glacial flip flop system is robust to change in the ice forming critical temperature parameter, even as this parameter goes through a time dependent change. Furthermore, the cycles exist even in the absence of the orbital forcing parameters, capturing the climate system's internal periodic behavior, akin to that exhibited by other conceptual models eg.  \cite{maasch1990low}. 

\begin{figure}
\centering

\subfloat[The phase portrait of the system, with orbital forcing from \cite{laskar2004long}, using ice forming temperature threshold $Tc_a$ (purple) from LR04 record's first SSA mode and a linearly increasing $Tc_r$ (pink). The LR04 record is from \cite{lisiecki2005pliocene}.]
{
\includegraphics[width=0.4\textwidth]{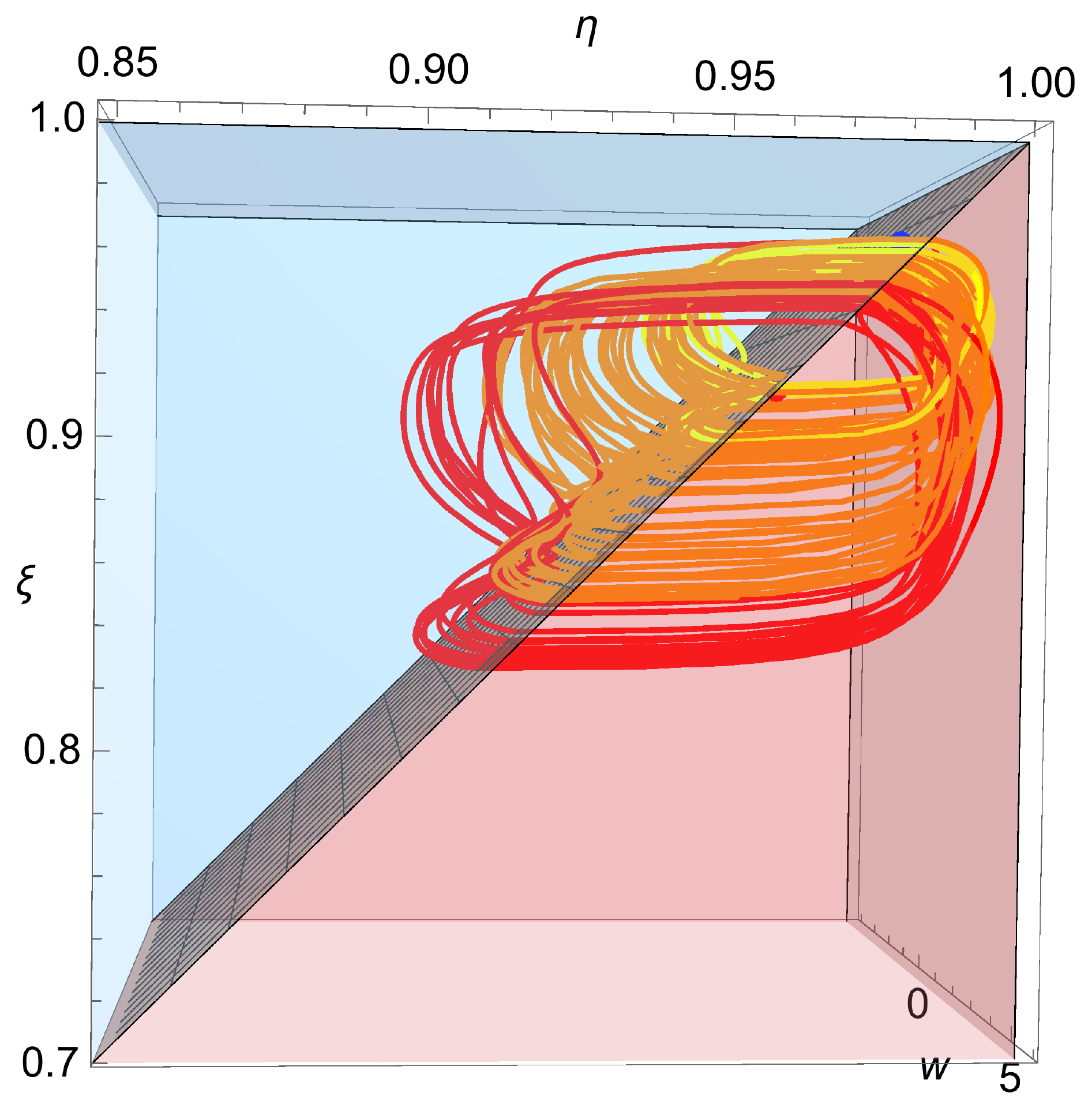}
\includegraphics[width=0.4\textwidth]{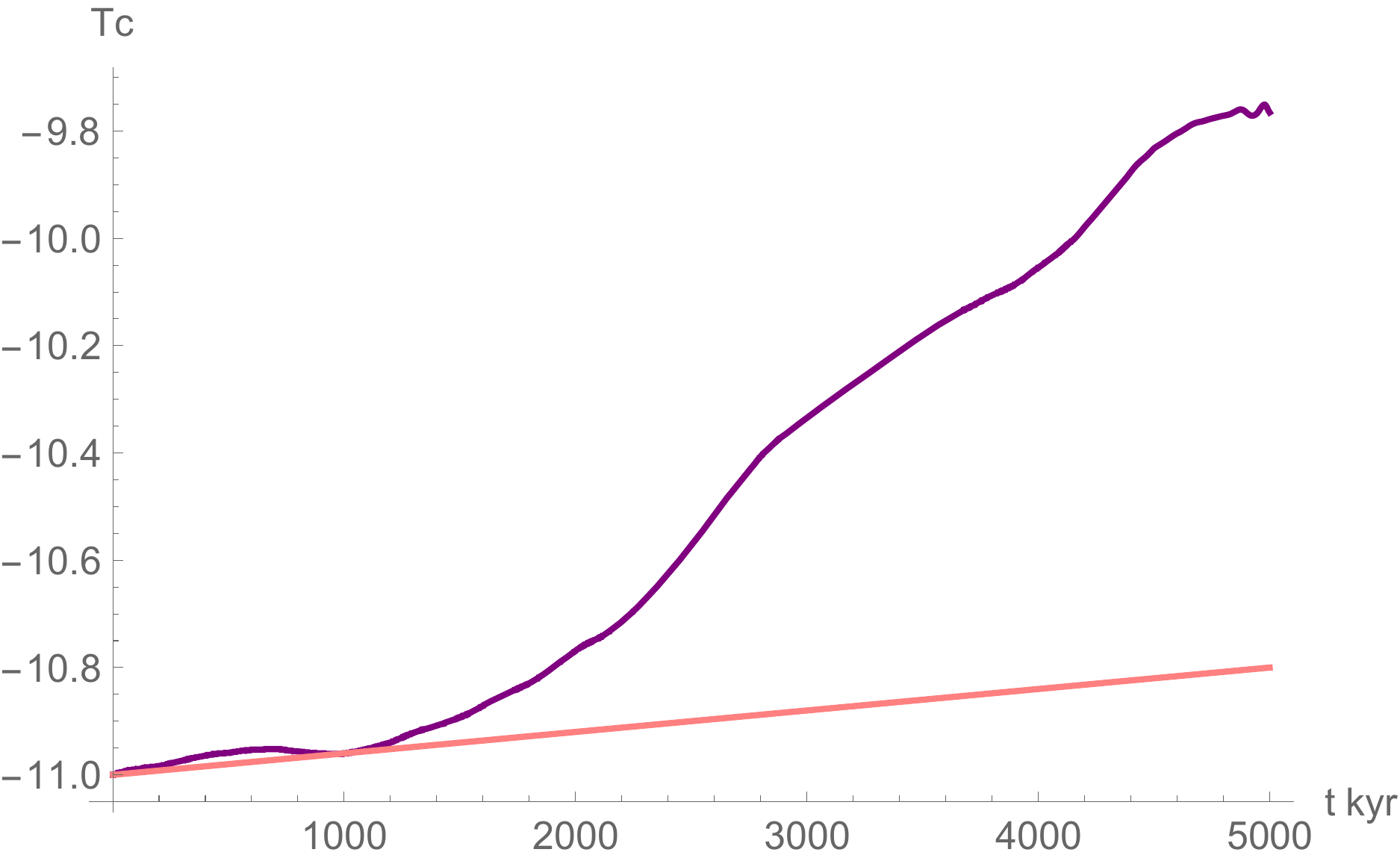}
}

\subfloat[The resulting lowest mode of the SSA reconstruction for both LR04 record and ice extent variable $\xi$ from the experiment using the ice forming temperature function above.]
{
\includegraphics[width=0.4\textwidth]{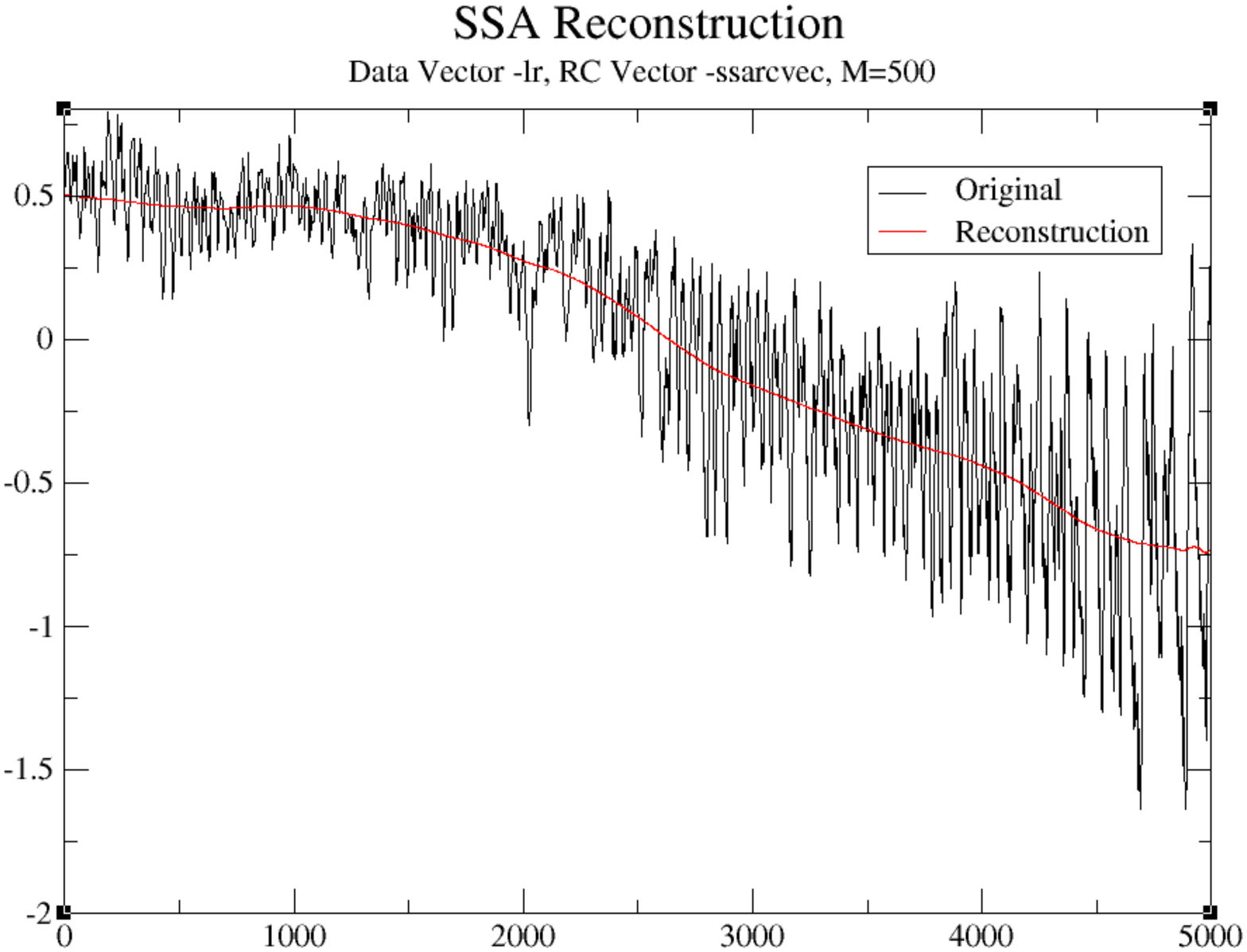}
\includegraphics[width=0.4\textwidth]{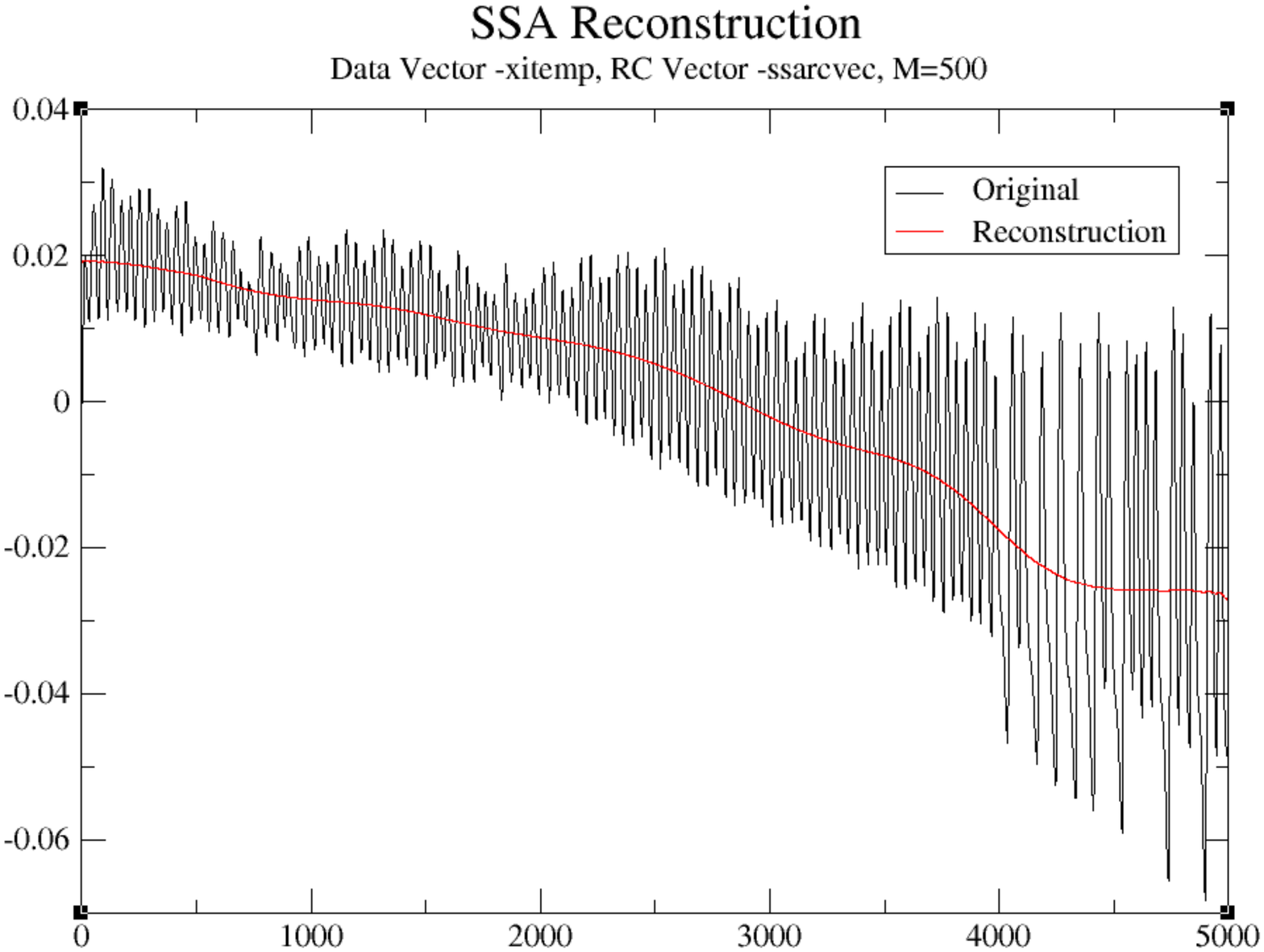}
}

\caption{Glacial flip flop, with orbital forcing and empirical ice forming critical temperature function.} \label{fig3}
\end{figure}

 The next exploration is aimed at obtaining a time series for the maximum ice extent $\xi$ so its period and amplitude over the past 5 Myr mimic that of the ice volume proxy $\delta^{18}$O record from \cite{lisiecki2005pliocene}. The results of are shown in Figure \ref{fig3}. To do this, eccentricity dependent parameter $Q$ and obliquity dependent $s_2$ are varied in time using the orbital data from \cite{laskar2004long}. The formulation for $Q$ and $s_2$ as functions of eccentricity and obliquity for the Budyko modeltakes a similar approach as  done in \cite{mcglehman2012}. Interestingly, the lowest mode of the singular spectrum analysis (SSA) method of ice volume proxy $\delta^{18}$O record from \cite{lisiecki2005pliocene} shows an increasing trend in ice volume. To explore further the response of the system to the nature of the increase in the ice forming temperature threshold, a linearly transformed first SSA mode of the $\delta^{18}$O from the past 5 Myr is used as an ice forming critical temperature threshold $Tc_a$. 

\section{Discussion}
A Budyko-Sellers type energy balance model is used as the basis a glacial cycle model called the \emph{glacial flip flop}. The conceptual model captures two important feedbacks: albedo and temperature-precipitation. A change in the ice forming critical temperature captures an MPT like phenomenon in the model. While the reasoning for the change in the critical temperature is beyond the scope of this work involved in developing the model, the model highlights some key non-linear feed backs that should be explored further. Furthermore, the latitudinally dependent variables developed in the model provide novel tools for further conceptual explorations of the Plio-Pleistocene glacial cycles. Mathematically, the non-smooth nature of the model is also interesting. One might wonder what would happen to the system if it is allowed to switch between the glacial and interglacial regimes.

\newpage

\clearpage

{\bf{\flushleft References}}
\bibliography{references.bib}

\end{document}